\documentclass[english,twoside,a4paper]{article}
\usepackage[latin1]{inputenc}
\usepackage[T1]{fontenc}
\usepackage{amsmath}
\usepackage{amsfonts}
\usepackage{graphicx}
\usepackage{lipsum}
\usepackage{trace}
\usepackage{subfigure}
\usepackage{a4wide}
\usepackage{amssymb}
\usepackage{fancyhdr}
\usepackage{mathrsfs}
\usepackage[toc,page]{appendix}
\usepackage{xcolor}
\usepackage{float}
\usepackage{lmodern}
\usepackage{hyperref}

\usepackage{lmodern}

\usepackage{tensor}
\usepackage{cite}
\newcommand{\be}{\begin{equation}}
\newcommand{\ee}{\end{equation}}
\newcommand{\bea}{\begin{eqnarray}}
\newcommand{\non}{\nonumber}
\newcommand{\eea}{\end{eqnarray}}

\begin{document}

\title{Topological regular black holes without Cauchy horizon}

\author{ 
Marco Calz\'a$^{1,2}$\footnote{E-mail address: marco.calza@unitn.it
},\,\,\,
Massimiliano Rinaldi$^{1,2}$\footnote{E-mail address: massimiliano.rinaldi@unitn.it
},\,\,\,
Sergio Zerbini $^1$\footnote{E-mail address: sergio.zerbini@unitn.it}
\\
\\
\begin{small}
$^1$ Department of Physics, University of Trento, Via Sommarive 14, 38123 Povo (TN), Italy
\end{small}\\
\begin{small}
$^2$ TIFPA-INFN,  Trento, Via Sommarive 14, 38123 Povo (TN), Italy
\end{small}\\
}

\date{}

\maketitle

\abstract{\noindent  
Regular and spherically symmetric black holes that solve the singularity problems of the Schwarzschild solution are phenomenologically viable at large distances but usually suffer from the Cauchy horizon instability. To overcome this drawback, we extended the analysis to include hyperbolic and toroidal horizon topologies within the framework of static, topologically maximally symmetric spacetimes. We show that both hyperbolic and toroidal black holes can be constructed without Cauchy horizons and without curvature singularities, thereby avoiding the mass inflation instability. These solutions exhibit asymptotic flatness in a generalized quasi-Minkowskian sense. 
The phenomenological aspects of these solutions are also studied by examining their thermodynamic properties, the photon sphere, and the effective potentials, ensuring consistency with observable properties such as black hole shadows. Lastly, we investigate a reconstruction technique within a scalar-tensor gravity framework, illustrating how the discussed metrics can arise from well-defined scalar field dynamics. Our investigation presents a viable pathway for constructing physically realistic, regular black holes in both General Relativity and modified gravity, broadening the landscape of singularity-free spacetimes and offering models that may better reflect the nature of strong gravitational fields in astrophysical and cosmological settings.
}

\section{Introduction}
\label{sec:introduction}

The remarkable imaging of black hole (BH) shadows by the Event Horizon Telescope (see Ref. \cite{EHT} and references therein) has revolutionized our understanding of astrophysical compact objects. These observations provide robust empirical evidence supporting the description of black holes with the Kerr metric, consistently with General Relativity (GR) and within a small margin of uncertainty. The Kerr black hole solution, as a vacuum solution of GR, serves as an elegant and widely accepted description of spinning compact objects that reduces to the Schwarzschild metric in the absence of rotation. However, its nature as an exact solution raises questions about whether it fully encapsulates the complexities of physical reality. Indeed, the existence of other compact objects with event horizons that deviate from the Kerr solution cannot yet be definitively ruled out.

\noindent This leads to an important conceptual challenge: although the Kerr solution is mathematically unique in GR, it suffers from physical limitations, most notably the presence of a central singularity \footnote{Recently, in addition to the singularity problem,  embedding static horizons in expanding space-times has also been shown to be problematic, see \cite{Faraoni:2024ghi}.}. This singularity, unavoidably arising from standard GR theorems, is nonphysical and probably signals the breakdown of classical physics in extreme regimes. Addressing this limitation requires exploring alternative models that can evade these singularities while preserving key physical properties, such as the existence of an event horizon. One promising approach involves regular black holes, which are characterized by the absence of singularities and are often associated with nonstandard matter fields. Alternatively, modified gravity frameworks allow for the construction of singularity-free black hole solutions even in the absence of matter, opening new avenues for exploration.

This study focuses on Static Topologically Maximally Symmetric (STMS) spacetimes, described by the general metric (see e.g. \cite{Geroch, Xantho,Vanzo,Birmingham,Mann:1997iz,Lemos,Aminneborg:1996iz,Mann:1996gj,Cai:1996eg,Brill:1997mf}):
\begin{equation}\label{general metric}
 ds^2=-f(r)dt^2+\frac{dr^2}{g(r)}+r^2(d\theta^2+h_k(\theta)^2 d\phi^2)\,,   
\end{equation}
where 
\begin{equation}\label{kdef}
    h_1(\theta)= \sin \theta\,, \quad h_0(\theta)=\theta\,, \quad  h_{-1}(\theta)= \sinh \theta\,. 
\end{equation}
In a more compact form, the above metric can be written as

\begin{equation}\label{topbh}
    ds^2=- f(r)dt^2+\frac{dr^2}{g(r)}+r^2d\theta^2+r^2\left[\sinh(\sqrt{-k}\,\theta)\over\sqrt{-k}\right]^2d\phi^2\,.\end{equation}

\noindent Thus, we are dealing with the spherical ($k=1$), toroidal ($k=0$) or hyperbolic ($k=-1$) topology of the $S^2$ sphere, $T^2$ torus, or compact hyperbolic
manifold $H^2$ respectively. 
These metrics embrace a set of space-times wider than the minimal natural starting point (the case with $k=1$) for exploring generalizations of BH solutions and are of particular interest in the context of regular black holes. Regular black holes have drawn attention in recent literature, both as toy models and as solutions inspired by quantum gravity principles ~\cite{Borde:1996df,AyonBeato:1998ub,AyonBeato:1999rg,Konoplich,Khlopov1,Bronnikov:2005gm,Berej:2006cc,Bronnikov:2012ch,Rinaldi:2012vy,Stuchlik:2014qja,Dymnikova,Schee:2015nua,Johannsen:2015pca,Myrzakulov:2015kda,Fan:2016hvf,Sebastiani:2016ras,Toshmatov:2017zpr,Chinaglia:2017uqd,Colleaux:2017ibe,Frolov:2017dwy,Bertipagani:2020awe,Nashed:2021pah,Simpson:2021dyo,Franzin:2022iai,Chataignier:2022yic,Ghosh:2022gka,Khodadi:2022dyi,Farrah:2023opk,Fontana:2023zqz,Boshkayev:2023rhr,Luongo:2023jyz,Luongo:2023aib,Cadoni:2023lum,Giambo:2023zmy,Cadoni:2023lqe,Luongo:2023xaw,Sajadi:2023ybm,Javed:2024wbc,Ditta:2024jrv,Al-Badawi:2024lvc,Ovgun:2024zmt,Corona:2024gth,Bueno:2024dgm,Konoplya:2024hfg,Pedrotti:2024znu,Bronnikov:2024izh,Kurmanov:2024hpn,Bolokhov:2024sdy,Agrawal:2024wwt,Belfiglio:2024wel,Stashko:2024wuq,Faraoni:2024ghi,Konoplya:2024lch,Khodadi:2024efq,Calza:2024qxn,Ansoldi:2008jw,Nicolini:2008aj,Sebastiani:2022wbz,Torres:2022twv,Lan:2023cvz,Calza:2022ioe,Calza:2024fzo,Calza:2024xdh,Ovalle}.

\noindent An important class of regular BH solutions with a spherical horizon topology exhibits the property \( f(r) = g(r) \), with \( g(r) \simeq 1 - A r^2 + \ldots \) at small \( r \), corresponding to an inner de Sitter core (dS). The presence of this dS core ensures the regularity of the spacetime at the center, resolving the singularity issue. For a review on regular black holes with $k=1$, see, for example, \cite{Sebastiani:2022wbz,Cadoni,Fontana:2023zqz}. For a general discussion concerning topological regular black holes, see \cite{Colleaux1,Colleaux2}.

In general, however, these solutions face a critical challenge: the emergence of an inner Cauchy horizon. The Cauchy horizon introduces a dynamical instability known as mass inflation \cite{Israel}, as well as kink instability \cite{Har}. Those instabilities are fundamentally related to the causal structure of spacetime and have been studied extensively \cite{ Bona, Herm,DiFilippo, DiFilippo2}. Similar issues have been analyzed in the context of higher-order derivative gravity theories, as explored in Ref. \cite{Berti}, and in relation to the correlation functions of the Hawking radiation in \cite{Fontana:2023zqz}.

Recent work \cite{Car} proposed a new perspective on the solution to the problem of mass inflation. Here, the authors assume a topologically spherical regular black hole with an inner de Sitter core, described by the asymptotically flat  metric \eqref{general metric} with
\begin{equation}
f(r)=g(r) = \frac{(r - r_H)(r - r_C)^3}{N(r)}\,,
\end{equation}
where \( r_C < r_H \), and \( N(r) \) is a suitably function ensuring asymptotic flatness. This model features two horizons. That is, an event horizon at \( r = r_H \) and a Cauchy horizon at \( r = r_C \). In particular, the Cauchy horizon is a triple root of \( g(r) \), leading to vanishing surface gravity at \( r = r_C \) which eliminates the exponential divergence responsible for mass inflation. Moreover, the triple root structure allows \( g(0) = 1 \), preserving the dS core.
This result bears conceptual similarities to the absence of mass inflation observed in certain higher-order gravity models (see Ref. \cite{Berti}) and highlights a new avenue for regular BH solutions by carefully controlling the inner causal structure.
 
Another possible approach to deal with the absence of the Cauchy horizon has been advocated in \cite{Casadio23, Casadio24}. Here, black hole solutions without a Cauchy horizon with spherical topology, have been studied. The behavior of the associated $f(r)$ was taken,  for  $r$ small, as  $f(r)\simeq1- 2 C r <0$. A simple  example is 
\begin{equation}
    f(r)=1-\frac{2 M}{r+A}\,,
\end{equation}
where $A>0$.
In this case, the price to pay is the divergence of the curvature invariants.
However, the singularity may be  considered weaker, in the sense that it is integrable (see  \cite{Estrada,Estrada1} for further considerations and references). Recently, this approach has been critically investigated  in \cite{Arrechea}.

In this paper, we propose a novel approach to regular black hole solutions that addresses the absence of Cauchy horizon instabilities. To achieve this, we extend the analysis to a broader class of space-times, allowing for more general topologies beyond the standard spherical case. By exploring these generalizations, we aim to provide a new perspective on constructing physically viable, regular black hole models within GR and modified gravity frameworks. With regard to this issue, 

The paper is organized as follows. In Sec. \ref{sec:TSSS&RBHS} we show that regular black holes with spherical horizon topology typically leads to the presence of Cauchy horizons. In Sec.\ \ref{overcoming} we show how to circumvent this sort of no-go theorem with a non-trivial horizon topology. In Sec. \ref{thermo} we study the first law of thermodynamics for these solutions and show that it holds provided we assume the standard GR equations. In Sec.\ref{sec:PhotonSphere} we study the properties of the photon sphere of these black hole metrics. In Sec. \ref{relax} we present a method to construct non-singular black holes with no Cauchy horizons but spherical horizon topology. In Sec. \ref{reconstruction} we propose a way to construct  a scalar-tensor effective action that yields, as solutions, the non-singular metrics that we discussed. We conclude with some remarks in Sec. \ref{sec:conclusions}.

\section{Regular black holes with general horizon topology}
\label{sec:TSSS&RBHS}

In this section we consider STMS metrics described by \eqref{general metric} and outline their main feature of interest for this work.
Let us, for instance, consider the subset of Eq.(\ref{general metric}) of the $tr$-symmetric metrics having $f(r)=g(r)$, where $r$ the areal radius. 

Often, it is useful to rewrite this metric in the form
\begin{equation}\label{general metric2}
 ds^2=-f(r)dt^2+\frac{dr^2}{f(r)}+r^2\left(\frac{d h_k^2}{1-k h_k^2}+h_k^2 d\phi^2\right)\,,   
\end{equation}
or, by the definition of 
\begin{align*}
    &x_1=2 \frac{h_k}{1+\dot{h}_k} \cos(\phi)\\
    &x_2=2 \frac{h_k}{1+\dot{h}_k} \sin(\phi)\;,
\end{align*}
where $\dot h= \partial_\theta h$,

\begin{equation}\label{general metric3}
 ds^2=-f(r)dt^2+\frac{dr^2}{f(r)}+\frac{r^2(dx_1^2+dx_2^2)}{\left[1+\frac{k}{4}(x_1^2+x^2_2)\right]^2}\,.
\end{equation}

\noindent The latter manifestly outlines that, in the case $k=0$, the angular part is flat. One may notice that this does not imply that, given $f(r) \rightarrow 1$ for $r\rightarrow \infty$, the whole metric is asymptotically Minkowskian. Indeed, the topology is different and we will refer to this behavior as quasi-Minkowskian.


In order to ensure the regularity of the solution, we check the three curvature invariants, which must be finite in the whole spacetime: the Ricci scalar $R$, the contraction of the Ricci tensor $R_{\mu \nu}R^{\mu \nu}$, and the Kretschmann $R_{\mu \nu \sigma \rho}R^{\mu \nu \sigma \rho}$. For the metric \eqref{general metric2} we find
\bea\label{33}
    R&=& -\frac{2 {\ddot h}_k+ r h_k (4 f'+ r f'') + 2 f h_k}{r^2 h_k}\,,\\\non
    R_{\mu \nu} R^{\mu \nu}&=& \frac{r^2 h^2_k ( 2 f' + r f'')^2 + 4 (f h + r h f' + {\ddot h}_k)}{2 r^4 h^2_k}\,,\\\non
    K&=&\frac{r^4 h_k^2 f''^2 + 4 r^2 h_k^2 f'^2 + 4 (f h_k + {\ddot h}_k)^2}{r^4 h_k^2}\,,
\eea
where $f'= \partial_r f$.
If use the metric written as in  \eqref{topbh}, we find
\bea\label{curvinv}
    R&=&-\frac{ r (4 f'+ r f'') + 2 (f-k) }{r^2 }\,,\\\non
    R_{\mu \nu} R^{\mu \nu}&=&  \frac{r^2  ( 2 f' + r f'')^2 + 4 (f -k  + r  f' )^2}{2 r^4}\,,\\\nonumber
    K&=&\frac{r^4  f''^2  + 4 r^2   f'^2 + 4 (f   - k )^2}{r^4 }\,.
\eea
Since ${\ddot h}_k=-k h_k$, expressions $\ref{33}$ and (\ref{curvinv}) coincide, as should be. 

It is possible to gain intuition on how should be a regular solution by inspecting these expressions. We ignore the so-called caustic solutions, in which the function $f$ has vertical asymptotes and discontinuities, and we only focus on the cases where $f$ is a continuous function of $r$. At first glance, there is no obvious general way to achieve finiteness for curvature invariants for all positive values of $r$, and it appears that the most critical point is $r=0$.
However, it is possible to highlight a simple common feature that allow to avoid divergences in $r=0$: that is, $f(0)-k$ must tend to 0 faster than $r^2$. Along these lines, we now show the possibility, provided by the topology, of avoiding both the singularity and the mass inflation problems.

 We close this Subsection observing that the fact  that the curvature invariant are finite might be not sufficient to ensure to deal with anon singular space-times. Mathematically one should investigate the geodesic completeness. Recently, in \cite{Rubio,Modesto}, this issue has been investigated for a large class of regular black holes having a de Sitter core. With regard to this, it is relevant to deal with metrics such as $f(r)=g(r)$ and  $f(-r)=f(r)$, (for example the well known Bardeen metric, defined in next Subsection ), since the geodesic completeness depends on the effective potential $V(r)$ related to a test particle. We will show in Section 5   that $V(r)$ depends on $f(r)$ also for space-time with  $k=0, -1$, namely also for a generic topological spherically symmetric space-time.
Now, $f(-r)=f(r)$ implies $V (-r) = V (r)$, which is relevant for the geodesic completeness. Having this in mind, nevertheless we will also continue to deal with metric like the Hayward one, (see the next Subsection) which does not satisfy this even property.

\subsection{A no-go theorem for the spherical topology}
\noindent As mentioned above, for $k=1$, one may cure the singularity at the origin modifying $f(r)$  in such a way that a de Sitter (dS) core is present and many proposals already exist, as discussed in \cite{Ansoldi:2008jw,Sebastiani:2022wbz,Cadoni,Fontana:2023zqz}. 
Two important examples  are the Bardeen regular BH  \cite{Bardeen}
\begin{equation}\label{Bardeen}
    f(r)=1-\frac{2m r^2}{(r^2+\ell^2)^{3/2}}\,,
\end{equation}
and the the Hayward regular BH \cite{Hayward}
\begin{equation}\label{Hayward}
    f(r)=1 - \frac{2 m r^2}{ r^3 + 2 m \ell^2}
\end{equation}
where $l$ may be considered  as a regularization parameter, which has dimension of a length. The singularity is absent because of the dS core: this means that, in the limit $r\rightarrow0$, both functions behave as
\be
f=1-\gamma r^2\,,
\ee
for some constant $\gamma$. Thus, the metric tends to be like the one of a de Sitter space near $r=0$.

However, to cure the central divergence with a dS core one must pay the price of the appearance of mass inflation instability associated with the presence of an inner Cauchy horizon. This is evident in both Bardeen's and Hayward's models, since $f$ has either none or two zeros (which may coincide in a very unstable configuration), the inner one being the location of the Cauchy horizon originating mass inflation.  For this class of regular BHs, this feature remains an open and serious issue.

Generally speaking, a function $f$, such that $f>0$ for $r>r_H$ ($r_H$ being its largest root) cannot ensure that the curvature scalars \eqref{curvinv} are finite in $r=0$, unless $(f(r)-1)\rightarrow 0 $ for $r \rightarrow 0$ faster than $r^2$. In this case, however, by continuity of $f$ there must be at least a second inner root $0<r_C<r_H$. By inspection it is clear that $f'(r_C)<0$ hence $r_C$ locates a inner Cauchy horizon, which causes mass inflation. The only exception is when the two roots coincide in a extremal configuration. In such a case, there is no Cauchy horizon but the black hole is physically unstable. Another option is to have more than two degenerate roots, such as in \cite{Car}.

As commented in the introduction, a possible way out is to invoke a triple (or higher) degenerate root of $f$ at the Cauchy horizon $r_C$ \cite{Car }, but despite this brilliant solution, there does not seem to be any other obvious way to avoid the singularity and the mass instability at the same time when $k=1$ and the metric is given by Eq.(\ref{general metric2}).

\noindent The same can be said if one takes into account Minkowski-core solutions such as the Culetu-Ghosh-Simpson-Visser BH \cite{Culetu,SVm}.

\noindent We can then formulate a no-go theorem for this case: \textit{In spherical topology a $tr$-symmetric metric can not avoid  inner horizon singularities and, at the same time, mass instability, unless the root of $f(r)$ at the Cauchy horizon has multiplicity larger than 2.}

\noindent In Sec.(\ref{relax}), we will comment on the consequences of relaxing the assumption on the metric and taking into account non-$tr$-symmetric metrics. 

\noindent In the next two Subsections we shall discuss how to avoid the mass inflation for regular BHs in the cases topological $tr$-symmetric metrics.


\section{Overcoming the no-go theorem with different horizon topologies}\label{overcoming}

In the above section we have shown that the presence of a Cauchy horizon and the absence of the central singularity are closely related. In particular, for spherical topology of the horizon one cannot simultaneously avoid Cauchy horizons and the central singularity. We found that the key to overcome this problem is to relax the requirement of spherical topology. We now present a study several examples along these lines.

\subsection{Hyperbolic non-singular black holes without Cauchy horizons}

In the following, we refer to the metric \eqref{topbh} with $g(r)=f(r)$. 
For the function $f(r)$  we explore two possibilities,
 one inspired by the Bardeen BH
\begin{equation}\label{Rmet}
  f(r)=1-{2\ell^2\over \ell^2+r^2}-{2mr^2\over(r^2+\ell^2)^{3/2}}\,,
\end{equation}
and the other by the Hayward BH
\begin{equation}\label{Rmet1}
  f(r)=1-{2\ell^2\over \ell^2+r^2}-{2mr^2\over r^3+2m\ell^2}\,.  
\end{equation}
In both cases, when $k=1$  and $k=0$ the invariants \eqref{curvinv} diverge for $r\rightarrow 0$. However, when $k=-1$ they vanish in the limit $r\rightarrow 0$ and for $\ell>0$, thus the black hole is regular.
In both cases, for $\ell \rightarrow 0$ the metric reduces to the Schwarzschild one, while for, $r \rightarrow 0$, one has
\begin{equation}
f(r)=-1+O(r^2)\,. 
\end{equation}
The core is therefore not of de Sitter type.
In addition, in both cases $f$ has only one root, which locates the event horizon $r_H$ and no inner Cauchy horizons are present.
Thus we have a regular and asymptotically flat BHs with Newtonian, but not Minkowkian behavior, without mass inflation problems.

\subsection{Toroidal non-singular black holes without Cauchy horizons}
A similar construction can be obtained for the  toroidal case, that is for the metric \eqref{topbh} with $k=0$. We choose the Bardeen-like and Hayward-like BHs function $f(r)$respectively as
\begin{equation}\label{torbar}
  f(r)=1-{\ell^2\over \ell^2+r^2}-{2mr^2\over(r^2+\ell^2)^{3/2}}\,,
\end{equation}
and 
\begin{equation}\label{torhay}
  f(r)=1-{\ell^2\over \ell^2+r^2}-{2mr^2\over r^3+2m\ell^2}\,. 
\end{equation}
As before, for both choices, when $k=1$ and $k=-1$ the curvature invariants diverge at the origin. However, for $k=0$,  these vanish at $r=0$. 

Both cases reduces to the Schwarzschild metric for a vanishing $\ell$, while, for $r \rightarrow 0$ and $\ell\neq 0$, one has
\begin{equation}
f(r)={(\ell-2m)r^2\over \ell^3}+O(r^4)\,,    
\end{equation}
for the Bardeen-like case and
\bea
f(r)=-{r^4\over \ell^4}+O(r^5)\,,
\eea 
for the Hayward-like case. The event horizon is unique and it is located at $r=2m$ for the metric \eqref{torhay} and at $r=\sqrt{4m^2-\ell^2}$ for the metric \eqref{torbar}. Thus, both metric represents regular and asymptotically flat BHs with Newtonian, but not Minkowkian behavior, without an inner Cauchy horizon.

For a regular toroidal black hole but  asymptotically AdS, see \cite{Fernandes:2025fnz}. Also in this case, the Cauchy horizon is absent.

\section{First Law of black holes thermodynamics}\label{thermo}
In this Section, we present an elementary derivation of the First Law for the kind of black hole metrics discussed above. The case with spherical horizon geometry can be found in \cite{Hayward94}. Here, we consider a static and spherically symmetric black hole described by the metric \eqref{topbh}.


It is important to stress that here we make the crucial assumption that the metric is derived from the standard Einstein equations with a source described by a matter stress tensor of the form $T_\mu^{\,\,\nu}={\rm diag}(-\rho, p_r, p_T,p_T)$. We then introduce coordinate-invariant quantities, which we will use in the following. The first is the so-called the Misner-Sharp (MS) energy. For a metric of the form
\be
ds^2=h_{\alpha\beta}dx^\alpha dx^\beta+r^2d\Omega^2\,,
\ee 
the MS energy is given by
\cite{Misner:1964je,Hayward94}
\be
E_{\rm MS}={r\over 2}\left(1-h^{\alpha\beta}\partial_\alpha r\partial_\beta r\right)\,
\ee 
Maeda et al. have generalized this expression for hyperbolic and toroidal geometries of the horizon \cite{Maeda:2006pm,Maeda:2007uu} to obtain
\be
E_{\rm MS}={r\over 2}\left(k-h^{\alpha\beta}\partial_\alpha r\partial_\beta r\right)\,.
\ee 
For the metric \eqref{topbh} this expression reduces to 
\begin{equation}\label{MSmass}
 E_{\rm MS}(r)=\frac{r}{2}\left[k-f(r)\right]\,.   
\end{equation}
On the horizon $r_H$, where  $f(r_H)=0$, one has
\begin{equation}
E_H=\frac{k r_H}{2}\,.
\label{ms}
\end{equation}
The second invariant quantity that we need is  the reduced normal trace of $T_{\mu}^{\,\,\nu}$ defined by
\begin{equation}
\tau=h_{\alpha\beta}T^{\alpha\beta}=T^{\,\,t}_t+T^{\,\,r}_r=-\rho+p_r\,,
\end{equation}
Since the Einstein tensor for the metric \eqref{topbh} has the property that $G_{r}^{\,\,r}=G_{t}^{\,\,t}$ then $p_r=-\rho$ and 
\begin{equation}
    \tau=-2\rho\,.
\end{equation}
Furthermore, the $tt$-component of the Einstein equations is
\begin{equation}
 8 \pi r^2 \rho=k-rf'-f \,.   
\end{equation}
By making use of this equation of motions, and evaluating all the quantities on the horizon (denoted the subscript $H$), one gets the following identity
\begin{equation}
k=r_Hf'_H-4 \pi \tau_H r^2_H\,.
\label{s}
\end{equation}
Now, since we are working within GR, the BH entropy satisfies the area law, hence 
\begin{equation}
 S_H=\frac{A_H}{4}=\frac{\alpha_k}{4} r^2_H\,, \quad A_H=\alpha_k r^2_H \,,   
\end{equation}
where $\alpha$ is a constant that depends on the chosen topology. For instance,   $k=1 $ implies $\alpha_1=4 \pi$.  As for the temperature of the BH, we assume that it coincides with the standard Hawking temperature, thus 
\begin{equation}\label{htemp}
    T_H=\frac{f'_H}{4 \pi}\,.
\end{equation}
Furthermore, we introduce the volume $V_H=v_k r^3_H$, where $v_k$ depends on the chosen topology of the horizon. For instance, $k=1$ implies $v_1=\frac{4 \pi}{3}$.  We now recall that the energy of the BH is the MS mass \eqref{MSmass} evaluated at the horizon, namely $E_k=\frac{ k r_H}{2}$. Then,  it is easy to show that the identity (\ref{s}) leads to
\begin{equation}
    dE_k=\frac{4\pi T_H }{\alpha_k} dS_H-\frac{2\pi \tau_H}{3v_k} d V_H\,.
\label{fl}
\end{equation}
This is the First Law of BH, where the pressure term is $P_H= \frac{2 \pi }{3 v_k}\tau_H$. For the case of spherical horizon ($k=1$) we have
\begin{equation}
    dE_1=T_H dS_H-\frac{\tau_H}{2} d V_H=T_H dS_H+\rho dV_H\,,
\end{equation}
where the second term vanishes in vacuum - that is the Schwarzschild case.

We conclude this Section observing that the identity (\ref{s}) may be used for deriving a modified Smarr formula. In fact, by combining (\ref{s}) and the definition (\ref{ms}), one finds
\begin{equation}
    E_H=2T_HS_H-\frac{3}{2}\tau_HV_H\,.
\end{equation}
In  vacuum, the above equation reduces to well-known standard Smarr formula.

\subsection{Hayward spherical black hole}
As an example, let us consider the Hayward BH with spherical topology, thus
\begin{equation}
    f(r)=g(r)=1-\frac{2m r^2}{r^3+2m l^2}=\frac{r^3+2 ml^2-2m r^2}{r^3+2m l^2}\,.
\end{equation}
We assume that $\ell$ is sufficiently small with respect to $m$ in order to have two horizons, the event horizon being the largest one located at $r=r_H$. Then $f(r_H)=0$, which implies
\begin{equation}
   r^3_H+2m l^2=2 m r^2_H\,.
\end{equation}
This equation can be inverted to yield
\begin{equation}
    m=\frac{r^3_H}{2(r_H^2-\ell^2)}\,,
\label{ho}
\end{equation}
so $m $ can be written as a function of the event horizon radius $r_H$. 
Now, if we assume standard GR, we have, as shown above, $p_r=-\rho$, and $\rho$ on the horizon reads
\begin{equation}
    \rho_H=\frac{3 \ell^2}{8 \pi r_H^4}\,.
\end{equation}
It follows that
\begin{equation}
    P_H=\frac{\tau_H}{2}= -\rho_H\,,
\end{equation}
and it is easy to see that the First Law (\ref{fl}) is satisfied as long as one identifies the  MS  mass evaluated at the horizon $M_{\rm MS}=\frac{r_H}{2}=$ with the energy of the BH.

It is clear that this result can be extended to all regular BHs with an inner dS core.

\subsection{ Toroidal Regular BH}
Let us now consider the more interesting case of the toroidal BH inspired by the Hayward metric \eqref{torhay}. The function $f(r)$ can be re-written in the form
\begin{equation}
    f(r)=\frac{ r^5}{(r^3+2 m \ell^2)(r^2+\ell^2)}(1-\frac{2 m}{r})=Q(r)(1-\frac{2m}{r})\,.
\end{equation}
where $Q(r)>0$ for $m>0$ and $\ell >0$ and, for large $r$, $Q(r) \longrightarrow 1$.
In such a way,  the existence of only one horizon at $r_H=2m$ becomes explicit. Then, the Hawking temperature \eqref{htemp} reads
\begin{equation}
    T_H=\frac{2m^3}{ \pi (4m^2+\ell^2)^2}\,.
\end{equation}
Again, the entropy is still given by the Area Law, as above. Since $k=0$, the
MS mass is vanishing, and the First Law becomes
\begin{equation}\label{k=0first}
    T_HdS_H={\alpha_0\tau_H\over 6v_0}dV_H\,.
\end{equation}

For the metric \eqref{torbar} the horizon is located at $r_H=\sqrt{4m^2-l^2}$ and the Hawking temperature is
\begin{equation}
    T_H=\frac{(4m^2-\ell^2)^{3/2}}{64 \pi m^4}\,.
\end{equation}
Also in this case, since $k=0$, the
MS mass is vanishing, and the First Law is \eqref{k=0first}.

\section{Photon Sphere}
\label{sec:PhotonSphere}

\noindent A serious phenomenological drawback of regular BH solution, and more in general of newly proposed BH solutions, might be the absence of photon rings or a radical modification of its shape in contrast with observations. For this reason in this section we compute the photon ring of the regular BHs without Cauchy horizons that we discussed above. The treatment is quite standard 
(see for example \cite{Perlick,Sunny22}) for spherical horizons. Here we shall present the case valid for any $k$, namely for a generic topological spherically symmetric space-times.

To start with, we recall the Lagrangian of a test particle in the metric \eqref{general metric}, it takes the form
\begin{equation}\label{lag}
 \mathcal{L}=\frac{\dot{s}^2}{2}=\frac{1}{2}\left[ -f(r)\dot{t}^2+\frac{\dot{ r}^2}{g(r)}+r^2\left(\dot{\theta}^2+ h_k^2(\theta) \dot{ \phi}^2  \right)\right]
\end{equation}
while the normalized four-velocity is
\begin{equation}
  g_{\mu \nu}\dot{x^\mu}\dot{x^\nu}=p  
\label{p}
\end{equation}
where a dot stands for derivative with respect to the affine parameter $\lambda$, and $p=0$ ($-1$) for massless (massive) particles. The symmetries of the metric \eqref{general metric} are encoded by the three Killing vectors $K^{\mu}=(1,0,0,0)$, $R^{\mu}=(0,0,0,1)$ $P^{\mu}=(0,0,1,0)$. The four-momentum conservation along the curves tangent to these vectors yield the conserved quantities
\begin{equation}\label{cons}
   E=K_{\mu}{dx^{\mu}\over d\tau}=-f(r)\dot{t}\,,\quad  L_{\phi}=R_{\mu}{dx^{\mu}\over d\tau}=r^2h_k^2( \theta) \dot{\phi}\,,\quad L_\theta=P_{\mu}{dx^{\mu}\over d\tau}=r^2\dot{\theta}\,,
\end{equation}
where we chose the affine parameter as the proper time $\tau$. The Euler-Lagrange equation for $\theta$ reads
\bea
r^2h_k{dh_k\over d\theta}\dot\phi^2=2r\dot r\dot\theta
+r^2\ddot\theta\,,
\eea
which can be written as
\begin{equation}
r^2h_k\frac{d h_k}{d \theta}  \dot{\phi}^2=  \frac{d}{d \tau}L_\theta\,.  
\end{equation}
The Euler-Lagrange equation for $\phi$ simply yields $L_\phi=$ constant. From these, it follows that the total (generalized) angular momentum 
   \begin{equation}
  L^2=L_\theta^2+\frac{L_\phi^2}{h_k^2( \theta)}\,,     
\end{equation}
is also conserved.  

Making use of \eqref{cons} and \eqref{p}, we find the first integral related to $r$, that is
\begin{equation}\label{firstint}
\dot{r}^2=g(r)\left( p+\frac{E^2}{f(r)}-\frac{L^2}{r^2}   \right)\equiv g(r)V(r)\,. 
\end{equation}
The effective Lagrangian can be written as
\bea
{\cal L}=\frac12\left({\dot r^2\over g(r)}+p-V(r)\right)\,,
\eea
so the Euler-Lagrange equation for $r$
\begin{equation}
    \frac{d}{d \lambda} \left(\frac{ \dot r}{g(r)} \right)=\frac{\partial {\cal L}}{\partial r}
\end{equation}
becomes
\begin{equation}\label{eomr}
    \frac{d}{d \lambda} \left(\frac{ \dot r}{g(r)} \right)=-\frac{V'}{2}-\frac{\dot r^2 g'}{2g^2}\,,
\end{equation}
where the prime stands for a derivative with respect to $r$.

The circular orbits are defined by the condition $\dot r_c=0$. Thus, \eqref{firstint} implies $V(r_c)=0$ while \eqref{eomr} leads to $ V'(r_c)=0$. These two constraints can be combined to yield
\begin{equation}
 E^2(r_cf'_c-2f_c)=2pf_c^2\,, \quad \frac{L^2}{E^2}=\frac{r_c^3 f_c'}{2f_c^2}=b^2\,,   
\end{equation}
where we have introduced the impact parameter $b=\frac{L}{E}$ and the notation $f(r_c)=f_c$.

In the case of mass-less particle  ($p=0$) these relations become
\begin{equation}
 (r_cf'_c-2f_c)=0\,, \quad b^2=\frac{r_c^2 }{f_c}\,.    
\label{pr}
\end{equation}
These two equations allows the computation of $r_c$ and $b$ for mass-less particle. For example, for the Scwarschild BH, one has  $f=1-\frac{r_s}{r}$,  $r_c=\frac{3r_s}{2}$, $b=\frac{3\sqrt{3}}{2}r_s$.

In the case of topological BHs in GR with a negative cosmological constant  (see, for example, \cite{Vanzo,Birmingham}) $g(r)=f(r)=k-\frac{r_s}{r}+ \frac{|\Lambda|}{3} r^2$ and

\begin{equation}
    2k=\frac{3 r_s}{r}\,.
\end{equation}
As a result, $k=0$ and $k=-1$ are not  allowed.  
 
\noindent However, in our models $f=g$ and the function $f$ can be written as $f=1-\frac{2M(r)}{r}$. Then,  \eqref{pr} can be rearranged as (here $M_c=M(r_c)$)
 \begin{equation}
     3M_c-r_cM'_c=r_c\,, \quad b^2=\frac{r_c^3 }{r_c - M_c}\,.
 \end{equation}
For the Bardeen choice, we have 
\begin{equation}
    M(r)=\frac{m r^3}{(r^2+\ell^2)^{3/2}}+\frac{(1-k) r\ell^2}{2(r^2+\ell^2)}
\end{equation}
Thus, for small $\ell$, for  $k=0$ and $k=-1$, one has 
\begin{equation}
 M(r)\simeq m+\ell^2\left( \frac{1-k}{2 r}-\frac{3 m }{2 r^2}  \right)\,.   
\end{equation}

\bea
r_c=3m-\frac16 {\ell^2\over m}(4k+1)+\ldots\,,
\eea 
which is a small correction to the  Schwarzschild case.  We note that in the same approximation, the event horizon  radius is

\begin{equation}
    r_H=2m-\frac{(1+2k)}{4m}\ell^2+...
\end{equation}
Similar conclusion may be drawn for the Hayward inspired metric.

For the sake of completeness,  we conclude this Section with the computation of the 3-dimensional tangential velocity related to circular orbits. Again, the computation is standard and can found in many papers.  Here we present the calculation for the sake of completeness.

In the case of general metric (\ref{M}, , one  can proceed as done above , and  one gets for spherical horizons and photon orbits

Thus
\begin{equation}
\dot{\rho}^2=g(\rho)\left( p+\frac{E^2}{f(\rho)}-\frac{L^2}{r(\rho)^2}   \right)\,,    
\end{equation}
and
\begin{equation}
    \dot{t}=-\frac{E}{f}\,,\quad \frac{L^2}{r^2}=r^2\dot{\Omega}^2\,,
\end{equation}
where
\begin{equation}
    \dot{\Omega}^2=\dot{\theta}^2+\sin^2 \theta \dot{\phi}^2\,.
\end{equation}
For circular orbits
\begin{equation}
\frac{L^2}{E^2}=\frac{r^3 f'}{2f^2}\,.
\end{equation}
Now, it is well known that the physical 3-dimensional velocity is given by

\begin{equation}
    v^i=\frac{d x^i}{ \sqrt{-g_{00}} d t}\,.
\end{equation}
As a result, for circular orbits

\begin{equation}
    v^2=\frac{r^2}{f}(\frac{d \Omega}{d t})^2=\frac{r^2}{f}\frac{\dot {\Omega}^2}{ \dot{t}^2}=\frac{f}{r^2}\frac{L^2}{E^2}\,.
\end{equation}
The final expression reads
\begin{equation}
  v^2=\frac{r(\rho)f'(\rho)}{2 f(\rho)}\,.  
\end{equation}
In the gauge where $r$ is the radial variable, this coincides with the Schwarzschild case, as expected.

\section{Regular black holes with no Cauchy horizons and spherical horizon topology}\label{relax}

In the previous Section, we have shown that one may have regular BHs without inner horizon but with hyperbolic or toroidal horizon geometry. We observe that this seems to happen when the radial coordinate $r$ is also the areal radius. 

Thus, inspired by an idea presented in \cite{Boos},  we work around this obstacle by considering black holes with spherical horizon but generalized areal radius. To wit, let us start from the following metric
\begin{equation}
    ds^2=-\left(1-\frac{2M(\rho)}{\rho}   \right)dt^2+
    \left(1-\frac{2M(\rho)}{\rho}   \right)^{-1}d\rho^2+
    r^2(\rho) dS^2\,,
\label{M}
\end{equation}
where, in the case of  Hayward-like regular BH,  $M(\rho$ ) is given by

\begin{equation}\label{R2met11h}
  M(\rho)= \frac{m \rho^3}{(\rho^3+2m\ell^2)}   >0\,.  
\end{equation}

and for Bardeen BH

\begin{equation}\label{R2met11b}
M(\rho)= \frac{m \rho^3}{(\rho^2+\ell^2)^{3/2}}  >0\,.  
\end{equation}
In the case of other regular BHs,  $M(\rho)$ may be constructed in the same way (see the Appendix).   


Now,  we make the following choice for the areal radius 
\begin{equation}
    r(\rho)=\left(\frac{m }{M(\rho)}\right)\rho\,,
\label{r}
\end{equation}
and, for the sake of simplicity, we consider the Hayward case only. Thus
\begin{equation}\label{areal}
    r(\rho)=\frac{\rho^3+2m\ell^2}{\rho^2}\,,
\end{equation}
and it clearly is a non-invertible function unless we restrict $\rho$. In fact $r(\rho)$ has an absolute minimum at $\rho=(4m\ell^2)^{1/3}$ and we choose the branch $\rho\geq (4m\ell^2)^{1/3}$, so that $r$ is an increasing function of $\rho$. Therefore, the range of $r$ is $[r_0,\infty[$ where $r_0^3=27m\ell^2/2$.
The curvature invariants for the metric \eqref{M} are all inversely proportional to powers of $\rho(\rho^3+2m\ell^2)$ thus they are singular at $\rho=0$. However, we will see that with the restriction obtained above on the range of $\rho$ the singularity is no longer physically present. 

The horizons are determined eventually by the usual equation $\nabla^\mu r(\rho)\nabla_\mu r(\rho)=0$ that is 
\begin{equation}
0=    \left(1-\frac{2M(\rho)}{\rho}   \right)\left(   \frac{d r(\rho)}{d \rho}\right)^2\,.
\label{h}
\end{equation}
One solution is the usual Hayward's, namely
\begin{equation}
    \rho_H^3+2m\ell^2=2m \rho_H\,.
\end{equation}
If $\rho$ is a real root, the horizon is located, in terms of $r$, at  $ r(\rho_H)=2m$. 
The other horizon is located where 
\begin{equation}
    0=\frac{d r(\rho)}{d \rho}= \frac{\rho^3-4m\ell^2}{\rho^3}\,,
\end{equation}
whose solution $\rho_\ell^3=4m \ell^2$ coincides with the minimum of the function \eqref{areal}. This corresponds to 
\begin{equation}
r(\rho_\ell)=\frac{6 m^{1/3}}{4^{2/3}}\ell^{2/3}\,.
\label{l}
\end{equation}
For $\ell << m$, one is dealing with a BH, with only the event horizon, since 
$f(\rho)=1-\frac{2 M(\rho)}{\rho}$ is not vanishing for $\rho_\ell$. Thus, the inner horizon is absent.
To clarify further these remarks, it is convenient  the use of Schwarzchild gauge, in which the areal radius coincides with the radial coordinate. Making use of equation (\ref{r}), one has
\bea
ds^2=-\left(1-{2m\over r}\right)dt^2+{M^2dr^2\over \left(1-{2m\over r}\right)(m-rM')^2}+r^2dS^2
\eea 
where $M=M(\rho(r))$, with $\rho(r)$ obtained inverting the relation (\ref{r}), and a prime denotes a derivative with respect to $\rho$.  In the case of Hayward the above metric becomes
\begin{equation}
 ds^2=-\left(1-\frac{2m }{r}   \right)dt^2 + 
{\rho^6 dr^2\over \left(1-{2m\over r}\right)(4m\ell^2-\rho^3)^2 }+ r^2dS^2\,,
\label{Mh}   
\end{equation}
where $\rho$ as a function of $r$ is implicitly determined by the relation \eqref{areal}, that is
\begin{equation}
 \rho^3-r \rho^2+2m\ell^2=0\,.   
\label{rho}
\end{equation}
In the Schwarzschild gauge, the curvature invariants diverge at $r=0$ and $\rho=0$. However,  the radial coordinate $r$ must lie in the interval $[r_0,\infty[$ (see above) thus this singularity is not physical. Similarly, we have also the condition $\rho^3\geq 4m\ell^2$ hence also the other singularity is not physical.
The horizons are located at the zeroes of $g^{rr}$. Thus, there is one at $r_H=2m$ provided $r_H>r_0$, which implies $4m\geq 3\sqrt{3}\ell$. The other horizon is located where $\rho^3=4m\ell^2$. However, the range of $\rho$ was chosen above to be $\rho^3\geq 4m\ell^2$. Therefore, what would be the inner horizon actually is a null surface that acts as the inner boundary for the spacetime described by the metric \eqref{Mh}. As such, it cannot be traversed and it is not a physical horizon.
A similar analysis can be carried on to the case of the Bardeen metric.

\section{Reconstruction method in a scalar-tensor model}\label{reconstruction}

Let us now investigate whether, within a simple scalar-tensor model, the metric (\ref{general metric}) may be obtained as an exact solution.  With regard to this issue, recently this task has been accomplished for the Hayward regular BH, but working within a scalar-Einstein-Gauss-Bonnet framework. \cite{Nojiri24}.
Here, we adopt a similar strategy so our starting point is the action
 \begin{equation}
  I=\int d^4x \sqrt{-g} \mathcal{L}=\int d^4x \sqrt{-g}\left( F(\phi)\frac{R}{2}-\frac{(\partial \phi)^2}{2}+V(\phi)  \right)\,,  
\end{equation}
where $F(\phi)>0$ is a suitable function to be determined. Given the metric \eqref{general metric} and the identity $d^2h_k(\theta)/d\theta^2=-k h_k(\theta)$ 
one has
\begin{align}
  &R=  \frac{2 \left(-r g'-g+\mathit{k}\right)-2 \sqrt{\frac{g}{f}} H'}{r^2}
\end{align}
where $H=\frac{1}{2} r^2 \sqrt{\frac{g}{f}} f'(r)$ and prime denoting the derivative with respect $r$.
Therefore,  for the gravitational Lagrangian, one has
\begin{equation}
    \sqrt{-g}\mathcal{L_{\rm grav}}= \sqrt{-g}\frac{R}{2} F(\phi)=\left(\sqrt{\frac{f}{g}}(k-g-r g')-H'\right) F(\phi) h_k\,.
\end{equation}
The matter Lagrangian reads
\begin{equation}
 \sqrt{-g} \mathcal{L_\phi}= \sqrt{-g}\left(-\frac{1}{2} g^{\mu \nu}\partial_\mu\phi \partial_v \phi+V(\phi)  \right)\,     
=-\frac{r^2 \sqrt{fg}}h_k{2}\phi'^2+r^2\sqrt{\frac{f}{g}}h_kV(\phi)\,,
\end{equation}
In both expression, the function $h_k(\theta)$ appears as a multiplicative factor hence it does not contribute to the equations of motion and we can omit it in the following. To work out the equations of motion it is convenient to integrate by parts the term $-H'F(\phi)$ in the gravitational Lagrangian by using the identity
\begin{equation}
    \frac{d (F H)}{dr }=F \frac{d H}{d r}+H \frac{d F}{d r}=F \frac{d H}{d r}+ H G \phi'\,,
\end{equation}
where we set $G=\frac{d F}{d \phi}$. Thus, we find the effective Lagrangian 
\bea\label{Lag}
\sqrt{-g}(\mathcal{L}_{\rm grav}+\mathcal{L}_\phi)=\sqrt{\frac{f}{g}}(k-g-rg')F(\phi)+\frac{r^2f' \sqrt{g}}{2 \sqrt{f}}G(\phi)\phi'-\frac{r^2 \sqrt{fg}}{2}\phi'^2+r^2\sqrt{\frac{f}{g}}V(\phi)\,,
\eea
where the variables are the functions $g,f,\phi$. 
We can now apply the usual variational principle by varying the above action with respect to $f,g,\phi$ and then by setting  $f=g$ as in  \cite{Nojiri24}. We find
\begin{align}
    &\delta f: \;\;\;\;\;\; g \frac{d}{d r } \left(F' r^2\right)= (k-g-r g') F - \frac{1}{2}g'F' r^2- \frac{1}{2} r^2 \phi'^2 g + r^2 V  \label{8} \\
    &\delta g: \;\;\;\;\;\; 2 g r F' = \left( k - g - r g' \right) F  - \frac{1}{2}  F' \, r^2 g' + \frac{1}{2} g r^2 \phi'^2 + r^2  V  \label{88}\\ 
    &\delta \phi: \;\;\;\;\;\;\frac{d}{dr}\left( \frac{G r^2 g'}{2} -r^2 g \phi' \right)= \left(k - g - r g'\right) G+
     \frac{dG}{d\phi} \frac{\phi' r^2 g'}{2} + r^2 \frac{dV}{d\phi} \label{888}
\end{align}
where we used the chain rule $G\phi'=F'$.

By subtracting (\ref{8}) and (\ref{88}) we find
\begin{equation}
F''=-\phi'^2\,,
\label{F}
\end{equation}
that is
\begin{equation}\label{phiF}
    \phi'=\pm \sqrt{-F''}\,,
\end{equation}
which implies $F''<0$, otherwise one is dealing with a ghost scalar field (negative kinetic energy).  Eq.(\ref{F}) is quite interesting, because it does not depend on any metric function and can be used to find $\phi(r)$, once $F(\phi)$ is known. In addition
  equations (\ref{8}) and (\ref{88}) converge to  the unique equation
\begin{equation}
  V=  \frac{g}{2}F''+\left(\frac{2g}{r}+ \frac{g'}{2}\right)F'-\left(\frac{k}{{r^2}}-\frac{g}{{r^2}}-\frac{g'}{r}\right)F\,.
\label{8888}
\end{equation}
Making use  of equation (\ref{F}), and taking the derivative with respect to $r$ ,
we have 
\begin{equation}
     F'''=-2\phi''\phi'\,. 
\label{F'''}
\end{equation}
Now, multiplying the equation   (\ref{888})  by $\phi'$, yields
\begin{equation}
    V'=\frac{gF'''}{2}+\left(   g'+\frac{2g}{r} \right)F''+\left(   \frac{4rg'+r^2g''-2k+2g}{2r^2}\right)F'\,.
\label{v'}
\end{equation}
By combining this equation with the derivative with respect to $r$ of \eqref{8888} finally gives a differential equation involving only $g$ and $F$, namely
\begin{equation}
    (2rg-r^2g')F'=(2k-2g+r^2g'')F\,.
\label{ff}
\end{equation}
This can be formally solved as
\begin{equation}
    F(r)=F_0 e^{\int dr \frac{2k-2g+r^2g''}{2rg-r^2g'}  }\,.
\label{Fr}
\end{equation}
where we take $F_0=1$ to guarantee $F(r)>0$.  
Given a metric function $g(r)$, one may obtain $F(r)$ by using \eqref{Fr}. Then, by invoking \eqref{phiF}, it is possible to compute $\phi(r)$, which, in principle, can be inverted to reconstruct $F(\phi)=F(r(\phi))$.

As first example, we take $k=1$ and $f=g$, with 
\begin{equation}
    g(r)=1+br-ar^2\,, \quad b >0,\,\, a>0\,.
\end{equation}
The metric represents a BH, which is a deformation of the de Sitter space (which correspondes to the case $b=0$). The horizon is located at
\begin{equation}
    r_H=\frac{b+\sqrt{b^2+4a }}{2}\,,
\end{equation}
and the radial coordinate is bounded by $0 < r < r_H  $. The exponent of \eqref{Fr} can be easily integrated and we find (with $F_0=1$)
\begin{equation}
  F(r)=\frac{1}{4(1+\frac{b}{2}r)^2}\,.  
\end{equation}
It is then immediate to verify that $F''(r)>0$ for all $r$ thus the scalar field is a ghost.  

In the case of de Sitter, $b=0$, $F=1/4$, and $\phi$ is a constant. In this case, the potential reads  $V=-3a$, namely  a cosmological constant term. It is easy to show that this result can be extended to the Schwarzschild-de Sitter case, namely 
\begin{equation}
    f(r)=g(r)=1-\frac{2m}{r}-ar^2\,,
\end{equation}
because, also for this metric one has   $  2-2g+r^2g''=0 $. Thus $F=1$ and $V=-3a $.

As a second non-trivial   example,  we consider the exact solution discussed in  \cite{Nadalini}. For the metric discussed here
\begin{equation}
g(r)=k\left( 1+\frac{r_0}{r} \right)^2+\frac{r^2}{l^2}\,.    
\end{equation}
We find
\begin{equation}
    F(r)=   \frac{r(r+2r_0)}{(r+r_0)^2}\,.
\label{f4}
\end{equation}
Since
\begin{equation}
    F''=-\frac{6r_0^2}{(r+r_0)^4} <0\,,
\end{equation}
then
\begin{equation}
    \phi'(r)=\pm \sqrt{6}\frac{r_0}{(r+r_0)^2}\,,
\end{equation}
and
\begin{equation}
    \phi(r)= \sqrt{6}\frac{r_0}{r+r_0}\,.
\end{equation}
This can be inverted to give
\begin{equation}
    r=\left(  \frac{\sqrt{6}}{\phi}-1  \right)r_0\,,
\end{equation}
and, by making use of (\ref{f4}), we find $F=1-\frac{\phi^2}{6}$,   and the potential
\begin{equation}
    V(\phi)={1\over \ell^2}\left(6+\frac{\phi^4}{12}\right)\,.
\end{equation}
This is in fact the  model, which describes  a scalar field  conformally coupled to gravity in presence of a  negative cosmological constant $\Lambda=-\frac{3}{l^2}$, discussed in \cite{Nadalini}.

What about the regular topological BHs metrics we have discussed in Section 3? For $k=0$, we observe that
\begin{equation}
 \frac{-2g+r^2g''}{2rg-r^2g'}= -\frac{d}{d r} \ln{(2rg-r^2g')} \,.
\label{Fr0}
\end{equation}
As a result
\begin{equation}
 F(r)=\frac{1}{2rg-r^2g'}\,.   
\end{equation}
Thus, in principle, one can explicitly compute $\phi(r)$, and   $F(\phi)$.

\section{Conclusions}
\label{sec:conclusions}
In this work, we have analyzed a broad class of regular black hole solutions characterized by a generalization of the horizon topology beyond the usual spherical case. Starting from the limitations of traditional regular BH models, particularly the inevitability of Cauchy horizons and the associated instabilities in spherically symmetric, tr-symmetric space-times, we have identified a no-go theorem that clarifies the impossibility of simultaneously avoiding singularities and mass inflation under such constraints.

To overcome these limitations, we extended the analysis to include hyperbolic and toroidal horizon topologies within the framework of static, topologically maximally symmetric spacetimes. Our findings demonstrate that both hyperbolic and toroidal regular black holes can be constructed without Cauchy horizons and without curvature singularities, thereby avoiding mass inflation instability. These solutions exhibit asymptotic flatness in a generalized, quasi-Minkowskian sense and satisfy the required curvature regularity conditions.

In the context of dynamically accreting black holes, it is shown in \cite{Carballo-Rubio:2024dca}, that mass inflation might arise even in the absence of Cauchy horizons. Whether such a mechanism applies to our topological non-singular black holes is an open question and deserves further investigation. Along the same lines, the relationship between the Strong Cosmic Censorship \cite{penrose} and the presence of Cauchy horizons seems to be absent in the present context since we consider non-singular black holes, thus circumventing the Conjecture independently of the horizon geometry. 

The asymptotic structure of these black holes may be of concern given their toroidal or hyperbolic section. In general, one has to consider that even if the Schwarzschild or Kerr black holes are asymptotically flat solutions, they are embedded in an expanding Universe, which has a different global set of spacetime symmetries \cite{Faraoni:2024ghi}. In such a case, one usually invokes the huge scale separation between local solutions and the large-scale structure of the Universe, and disregards the problem of matching the two solutions (say, e.g., Schwarzschild and Robertson-Walker) at some large distance. Therefore, one may advocate a similar argument and conclude that the local topology of the horizon is not relevant at cosmological distances. Moreover, at very large distances, and independently of the horizon topology, an observer far away from such BHs would not be able to locally measure the manifold topology. From the observational standpoint, we have shown that the photon sphere changes marginally compared to the usual Schwarzschild solution. Nevertheless, toroidal black holes (singular or not) can be of interest as they might form during a merger of binary black holes, see e.g. \cite{teut,kaplan}, or be responsible for quasar emission \cite{Spivey:2000rm}. 

Furthermore, we have shown that a novel metric Ansatz employing a generalized areal radius allows for the construction of regular BHs with spherical horizons that are free of inner horizons. This approach circumvents the restrictions of the no-go theorem by modifying the geometric interpretation of the radial coordinate.

In addition to the spacetime geometry, we explored the thermodynamic consistency of these solutions by deriving a generalized first law of black hole thermodynamics applicable to various horizon topologies. We verified the law's validity through specific examples, including both Hayward-like and Bardeen-like models, highlighting the role of the Misner-Sharp mass and the pressure-like contributions from the matter content.

We also addressed the phenomenological aspects of the proposed solutions by examining their photon spheres and effective potentials, ensuring consistency with observable properties such as BH shadows. Lastly, we provided a reconstruction technique within a scalar-tensor gravity framework, illustrating how the discussed metrics can arise from well-defined scalar field dynamics.

Overall, our study presents a viable pathway for constructing physically realistic, regular black holes in both General Relativity and modified gravity, broadening the landscape of singularity-free spacetimes and offering models that may better reflect the nature of strong gravitational fields in astrophysical and cosmological settings.

\section*{Acknowledgments}
\noindent We acknowledge support from the Istituto Nazionale di Fisica Nucleare (INFN) through the Commissione Scientifica Nazionale 4 (CSN4) Iniziativa Specifica ``Quantum Fields in Gravity, Cosmology and Black Holes'' (FLAG). M.C. acknowledges support from the University of Trento and the Provincia Autonoma di Trento (PAT, Autonomous Province of Trento) through the UniTrento Internal Call for Research 2023 grant ``Searching for Dark Energy off the beaten track'' (DARKTRACK, grant agreement no.\ E63C22000500003).

\appendix

\section{Generic Regular Black Holes with a de Sitter core}

In this Appendix,  we briefly  report the general approach to dealing with the regular spherical BH class having the de Sitter core. Here we follow   \cite{Sebastiani:2022wbz}. 

First, let us denote by  $h(x)$ a summable smooth function such that

\begin{equation}
   4\pi\int^\infty_0 x^2 h(x)dx=C < \infty\,.
\end{equation}
and consider
\begin{equation}
  f(r)=1-\frac{2M(r)}{r}=1-r^2G(r)\,.  
\end{equation}
The mass function related to the function $G(r)$ is $2M(r)=r^3G(r)$. The new function $G(r)$ depend on the function    $h(r)$ , and   is defined by      
\begin{equation}
 G(r)=\frac{8\pi m}{r^3 C}\int_0^{r/\ell}dy y^2 h(y)\,,   
\end{equation}
where $m$ is a mass parameter, and $\ell$ a suitable very small parameter.

Let us study the properties of this function $G(r)$. For large $r$ , one has $G(r)\simeq \frac{2 m}{r^3}+...$. Thus

\begin{equation}
    f(r) \simeq 1-\frac{2m}{r}+..\,.
\end{equation}
As a consequence, $m$ is the BH mass. 

For $r$ small, since we are assuming $h(r)$ smooth,  with 
$h(0)=1$,    namely

\begin{equation}
h(r)\simeq 1+h_1r+h_2r^2+...\,,
\label{c}
\end{equation}
 one has 

\begin{equation}
G(r)\simeq \frac{8\pi m}{3C\ell^3}+ \frac{8\pi m h_1 r}{4C\ell^4}+..,.
\label{e2}
\end{equation}
As a result, 
\begin{equation}
   f(r)=1- \frac{8\pi m}{3C\ell^3}r^2+...=1-\frac{\gamma}{3\ell^3} r^2+..\,,
\label{e1}
\end{equation}
  where $ \gamma=\frac{8\pi m}{C}  $ ,  namely a de Sitter core is present.  

In general, if the mass is sufficiently large,  $f(r)$ becomes negative, and two horizons are present: the metric corresponds to a BH with event horizon and a Cauchy horizon.  If the mass is sufficiently small,    $f(r)>0$,   one may deal with a compact horizonless object.

In fact, we may rewrite

\begin{equation}
 f(r)=1-\frac{8\pi m}{r C}\int_0^{r/\ell}dy y^2 h(y)\,.   
\end{equation}
The local minimum may be determined by  $f'=0$. Thus

by      
\begin{equation}
\int_0^{r_0/\ell}dy y^2 h(y)=\frac{r_0^3}{\ell^3}h(\frac{r_0}{\ell})\,,   
\label{min}
\end{equation}
which permits to compute $r_0$.  In fact, one has  $r_0=A\ell$ with

\begin{equation}
\int_0^{A}dy y^2 h(y)=\frac{A^3}{\ell^3}h(A)\,,   
\label{mina}
\end{equation}

Making use of the  equation (\ref{min}), one has

\begin{equation}
    f(r_0)=1-\frac{8\pi m r_0^2}{C \ell^3}h(\frac{r_0}{\ell})=1-  \frac{8\pi A^2 h(A)}{C} \frac{m}{\ell}\,.
\label{min1}
\end{equation}
As a consequence,   one has a horizonless compact object if 
\begin{equation}
    \ell >\frac{8\pi  A^2 h(A)}{C } m\,.
\end{equation}
And one is dealing with a BH, as soon as 

\begin{equation}
    \ell <\frac{8\pi A^2 h(A)}{C } m\,.
\end{equation}
As a simple example, we may consider the Fan-Wang BH \cite{Fan}.
In this case   $h(y)=\frac{3}{(1+y)^4}$. The constant $C=4\pi$. The related lapse function
\begin{equation}
    f(r)=1-\frac{2 m r^2}{(r+\ell)^3}\,.
\end{equation}
The location of the local minimum may be computed by making use of equation (\ref{mina}). The result is $r_0=2\ell$, and equation (\ref{min1}) gives
\begin{equation}
 f(r_0)=1- \frac{8}{27} \frac{ m}{ \ell}\,.   
\end{equation}
The other two examples used in the text can be treated in the same way.   First,  if we choose

\begin{equation}
h(y)=\frac{1}{(1+y^2)^{5/2}}\,, \quad C=\frac{4 \pi }{3}\,.
\end{equation}
The related $G(r)$
\begin{equation}
 G(r)=\frac{2 m}{(r^2+\ell^2)^{3/2}}\,,   
\end{equation}
and the associated  metric is the Bardeen metric \cite{Bardeen}, namely 

\begin{equation}
f(r)=1-\frac{2 m r^2}{(r^2+\ell^2)^{3/2}}\,.
\end{equation}
The  second example is related to the choice 
\begin{equation}
   h(y)=\frac{1}{(1+y^3)^{2}}\,, \quad C=\frac{4 \pi }{3}\,. 
\end{equation}
Thus
\begin{equation}
 G(r)=\frac{2 m}{(r^3+\ell^3)}\,,   
\end{equation}
and the associated  metric is the Hayward-like  metric \cite{Hayward}, namely 

\begin{equation}
f(r)=1-\frac{2 m r^2}{(r^3+\ell^3)}\,.
\end{equation}

Furthermore, also the discussion of Section 3  can be generalized. In fact, given the generic $G(r)$,  we may define another generic mass function according to
\begin{equation}
f(r)=1-\frac{(1-k)\ell^2 }{ r^2+\ell^2}-r^2 G(r)\,,
\end{equation}
valid only for $k=-1$ and $k=0$.
For small $r$, one has
\begin{equation}
f(r)=k+(\frac{1-k}{\ell^2}-\gamma)r^2+...   \,.
\end{equation}
The Cauchy horizon is not present,  and $f(0)=k$.

Also the discussion presented in Section 6 can be generalized making use of $2M(\rho)=\rho^3 G(\rho)$.
  

\end{document}